\documentclass[12pt]{article}
\usepackage{epsfig}
\newcommand{\bea}{\begin{eqnarray}}
\newcommand{\ena}{\end{eqnarray}}

\begin{document}
\baselineskip=15pt
\begin{titlepage}
\setcounter{page}{0}

\begin{flushright}
OU-HET 529 \\
astro-ph/0505253
\end{flushright}

\begin{center}
\vspace*{5mm}
{\Large \bf Parametrization of Quintessence and Its Potential}\\
\vspace{15mm}

{\large
Zong-Kuan Guo,$^{a}$
\footnote{e-mail address: guozk@itp.ac.cn}
Nobuyoshi Ohta$^b$
\footnote{e-mail address: ohta@phys.sci.osaka-u.ac.jp}
and
Yuan-Zhong Zhang$^{c,a}$}\\
\vspace{10mm}
\it
$^a$Institute of Theoretical Physics, Chinese Academy of
   Sciences, P.O. Box 2735, Beijing 100080, China \\
$^b$Department of Physics, Osaka University, Toyonaka,
   Osaka 560-0043, Japan \\
$^c$CCAST (World Lab.), P.O. Box 8730, Beijing 100080, China
\end{center}

\vspace{20mm}
\centerline{\large \bf Abstract}
{We develop a theoretical method of constructing the
quintessence potential directly from the effective equation
of state function $w(z)$, which describes the properties of
the dark energy.
We apply our method to four parametrizations of equation of
state parameter and discuss the general features of the
resulting potentials.
In particular, it is shown that the constructed quintessence
potentials are all in the form of a runaway type.}
\vspace{2mm}

\begin{flushleft}
PACS number(s): 98.80.Cq, 98.65.Dx
\end{flushleft}

\end{titlepage}


Recent observations of type Ia supernovae suggest that the
expansion of the universe is accelerating and that two-thirds
of the total energy density exists in a dark energy component
with negative pressure~\cite{AGRSP,SP}.
In addition, measurements of the cosmic microwave
background~\cite{DNS} and the galaxy power spectrum~\cite{MT}
also indicate the existence of the dark energy.
The simplest candidate for the dark energy is a cosmological
constant $\Lambda$, which has pressure
$P_\Lambda=-\rho_\Lambda$. Specifically, a reliable model
should explain why the present amount of the dark
energy is so small compared with the fundamental scale
(fine-tuning problem) and why it is comparable with the
critical density today (coincidence problem).
The cosmological constant suffers from both these problems.
One possible approach to constructing a viable model for
dark energy is to associate it with a slowly evolving and
spatially homogeneous scalar field $\phi$, called
``quintessence''~\cite{RP}-\cite{PJS}. Such a model
for a broad class of potentials can give the energy density
converging to its present value for a wide set of initial
conditions in the past and possess tracker behavior.

The dark energy is characterized by its equation of state
parameter $w$, which is in general a function of redshift $z$
in quintessence models.
The quintessence potential $V(\phi)$ and the equation of state
$w_\phi(z)$ may be reconstructed from supernova
observations~\cite{DHMT,AAS,TC}.
In this letter we develop a theoretical method of constructing
the quintessence potential $V(\phi)$ directly from the dark
energy equation of state function $w_\phi(z)$.
We apply this method to four typical parametrizations which
fit the data well~\cite{SHEM}-\cite{BFGGE} and discuss
the general features of the resulting potentials. The typical
behavior of the constructed potentials is found to be a
runaway type.


We consider a spatially flat FRW universe which is dominated
by the non-relativistic matter and a spatially homogeneous
scalar field $\phi$. The Friedmann equation can be written as
\begin{equation}
\label{friedmann}
H^2=\frac{1}{3M_{pl}^2}(\rho_m+\rho_\phi),
\end{equation}
where $M_{pl} \equiv (8\pi G)^{-1/2}$ is the reduced Planck
mass and $\rho_m$ is the matter density. The energy density
$\rho_\phi$ and pressure $P_\phi$ of the evolving scalar field
$\phi$ are given by
\begin{eqnarray}
\label{rp}
\rho_{\phi} &=& \frac{1}{2}\dot{\phi}^2+V(\phi), \\
P_{\phi}    &=& \frac{1}{2}\dot{\phi}^2-V(\phi),
\label{pp}
\end{eqnarray}
respectively, where $V(\phi)$ is the scalar field potential.
The corresponding equation of state parameter is now given by
\begin{equation}
\label{esp}
w_{\phi} \equiv \frac{P_{\phi}}{\rho_{\phi}}.
\end{equation}
Using Eqs.~(\ref{rp}) and (\ref{pp}), we have
\begin{eqnarray}
\label{dphi}
\frac{1}{2}\dot{\phi}^2 &=& \frac{1}{2}(1+w_{\phi})\rho_{\phi},\\
V(\phi) &=& \frac{1}{2}(1-w_{\phi})\rho_{\phi}.
\label{vphi}
\end{eqnarray}
The evolution of quintessence field is governed by the
equation of motion
\begin{equation}
\label{emp}
\dot{\rho_\phi}+3H(\rho_\phi+P_\phi)=0,
\end{equation}
which yields
\begin{eqnarray}
\rho_\phi (z) &=& \rho_{\phi 0}
 \,\exp \left[3\int_{0}^{z}(1+w_\phi)d\ln (1+z)\right]
 \nonumber \\
 & \equiv & \rho_{\phi 0}\,E(z),
\label{rhoz}
\end{eqnarray}
where $z$ is the redshift which is given by $1+z = a_0/a$ and
subscript $0$ denotes the value of a quantity at the redshift
$z=0$ (present).
In terms of $w_{\phi}(z)$, the scalar field potential
(\ref{vphi}) can be written as a function of the redshift $z$:
\begin{equation}
\label{vz}
V[\phi(z)]=\frac{1}{2}(1-w_{\phi})\rho_{\phi 0} E(z).
\end{equation}
With the help of $\rho_m=\rho_{m0}(1+z)^3$ and
Eq.~(\ref{rhoz}), the Friedmann equation (\ref{friedmann})
becomes
\begin{equation}
\label{fz}
H(z)=H_0\left[\Omega_{m0}(1+z)^3+\Omega_{\phi 0}
 E(z)\right]^{1/2},
\end{equation}
where $\Omega_{m0} \equiv \rho_{m0}/(3 M_{pl}^2 H_0^2)$
and $\Omega_{\phi0} \equiv \rho_{\phi0}/(3 M_{pl}^2 H_0^2)$.
Using Eq.~(\ref{dphi}), we have
\begin{equation}
\label{dphiz}
\frac{d\phi}{dz}=\mp
 \frac{(1+w_\phi)^{1/2}}{(1+z)H(z)}
 \left[\rho_{\phi}(z)\right]^{1/2},
\end{equation}
where the upper (lower) sign applies if $\dot{\phi}>0$
($\dot{\phi}<0$). The sign in fact is arbitrary, as it can be
changed by the field redefinition, $\phi \to -\phi$.
So we choose the upper sign in the following discussions.
Substituting Eqs.~(\ref{rhoz}) and (\ref{fz}) into
Eq.~(\ref{dphiz}) gives
\begin{equation}
\label{dpz}
\frac{d\phi}{dz} = - \sqrt{3}M_{pl}
 \frac{(1+w_\phi)^{1/2}}{(1+z)}
 \left[1+r_0 (1+z)^3 E^{-1}(z)\right]^{-1/2},
\end{equation}
where $r_0 \equiv \Omega_{m0}/\Omega_{\phi 0}$ is the energy
density ratio of matter to quintessence at present time.

We define dimensionless quantities
\begin{equation}
\tilde{V} \equiv V / \rho_{\phi 0}, \quad
\tilde{\phi} \equiv \phi / M_{pl}.
\end{equation}
The construction equations (\ref{vz}) and (\ref{dpz}) can
then be written as
\begin{eqnarray}
\label{TVZ}
\tilde{V}[\phi(z)] &=& \frac{1}{2}(1-w_{\phi}) E(z),\\
\frac{d\tilde{\phi}}{dz} &=& - \sqrt{3} \,
 \frac{(1+w_\phi)^{1/2}}{(1+z)}
 \left[1+r_0 (1+z)^3 E^{-1}(z)\right]^{-1/2},
\label{TPZ}
\end{eqnarray}
which relate the quintessence potential $V(\phi)$ to the
equation of state function $w_\phi(z)$.
Given an effective equation of state function $w_\phi(z)$,
the construction equations (\ref{TVZ}) and (\ref{TPZ}) will
allow us to construct the quintessence potential $V(\phi)$.

Our method is new in that it relates directly the
quintessence potential to the equation of state function, and
so enables us to construct easily the potential without
assuming its form. For instance, in the reconstruction method
discussed in Ref.~\cite{DHMT}, the reconstruction equations
relate the potential and the equation of state to
measurements of the luminosity distance. The potential may
thus be reconstructed by way of the luminosity distance from
supernova data. Usually this can be done by assuming the
form of a potential $V(\phi)$.
The dark energy properties are well described by the
effective equation of state parameter $w_\phi(z)$ which in
general depends on the redshift $z$.

Let us now consider the following
four cases~\cite{SHEM}-\cite{BFGGE}: a constant
equation of state parameter and three two-parameter parametrizations.

{\bf Case I}: $w_\phi=w_0$ (Ref.~\cite{SHEM})
\begin{eqnarray}
\label{pot1}
\tilde{V}(z) &=& \frac{1}{2}(1-w_0)(1+z)^{3(1+w_0)},\\
\frac{d\tilde{\phi}}{dz} &=& - \sqrt{3} \,
 \frac{(1+w_0)^{1/2}}{(1+z)}
 \left[1+r_0 (1+z)^{-3w_0}\right]^{-1/2}.
\label{phi1}
\end{eqnarray}

{\bf Case II}: $w_\phi=w_0+w_1 z$ (Ref.~\cite{ARCDH})
\begin{eqnarray}
\tilde{V}(z) &=& \frac{1}{2}(1-w_0-w_1 z)
 (1+z)^{3(1+w_0-w_1)}e^{3w_1 z},\\
\frac{d\tilde{\phi}}{dz} &=& - \sqrt{3} \,
 \frac{(1+w_0+w_1 z)^{1/2}}{(1+z)}
 \left[1+r_0 (1+z)^{-3(w_0-w_1)}
 e^{-3w_1 z}\right]^{-1/2}.
\end{eqnarray}

{\bf Case III}: $w_\phi=w_0+w_1\frac{z}{1+z}$
(Ref.~\cite{CP,EVL,TP})
\begin{eqnarray}
\tilde{V}(z) &=& \frac{1}{2}
 \left(1-w_0-w_1\frac{z}{1+z}\right)
 (1+z)^{3(1+w_0+w_1)}e^{-3w_1 \frac{z}{1+z}},\\
\frac{d\tilde{\phi}}{dz} &=& - \sqrt{3} \,
 \frac{(1+w_0+w_1 \frac{z}{1+z})^{1/2}}{(1+z)}
 \left[1+r_0 (1+z)^{-3(w_0+w_1)}
 e^{3w_1 \frac{z}{1+z}}\right]^{-1/2}.
\end{eqnarray}

{\bf Case IV}: $w_\phi=w_0+w_1\ln(1+z)$ (Ref.~\cite{BFGGE})
\begin{eqnarray}
\tilde{V}(z) &=& \frac{1}{2}
 \left[1-w_0-w_1\ln(1+z)\right]
 (1+z)^{3(1+w_0)+\frac{3}{2}w_1\ln(1+z)},\\
\frac{d\tilde{\phi}}{dz} &=& - \sqrt{3} \,
 \frac{\left[1+w_0+w_1\ln(1+z)\right]^{1/2}}{(1+z)}
 \left[1+r_0 (1+z)^{-3w_0-\frac{3}{2}w_1\ln(1+z)}
 \right]^{-1/2}.
\end{eqnarray}

\begin{figure}
\begin{center}
\includegraphics[width=11cm]{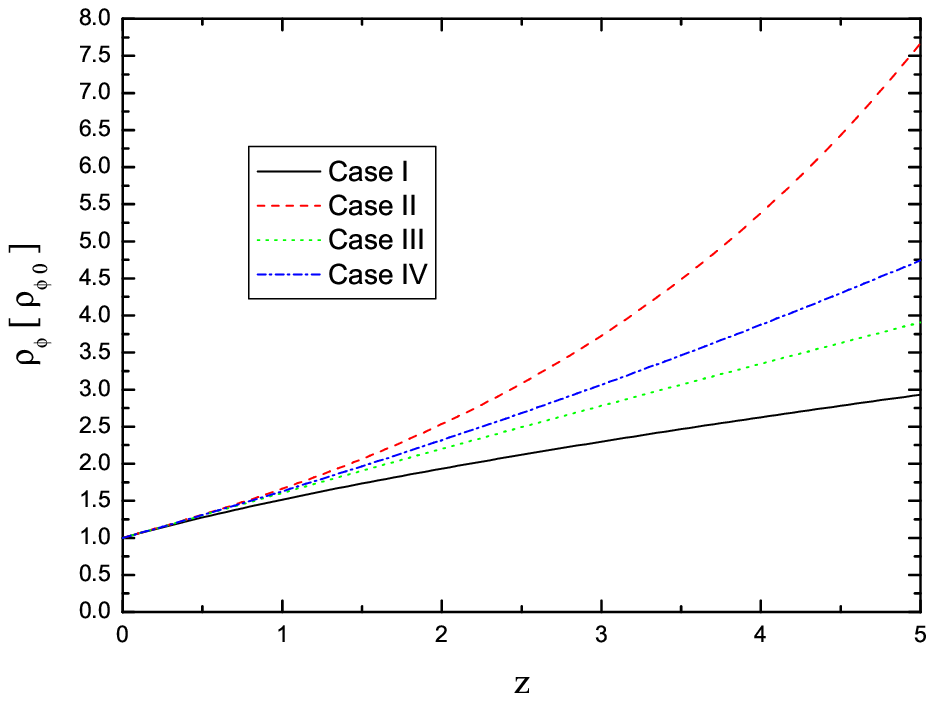}
\caption{Evolution of the energy density of the
quintessence $\rho_\phi (z)$.}
\label{fig:rhoz}
\end{center}
\begin{center}
\includegraphics[width=11cm]{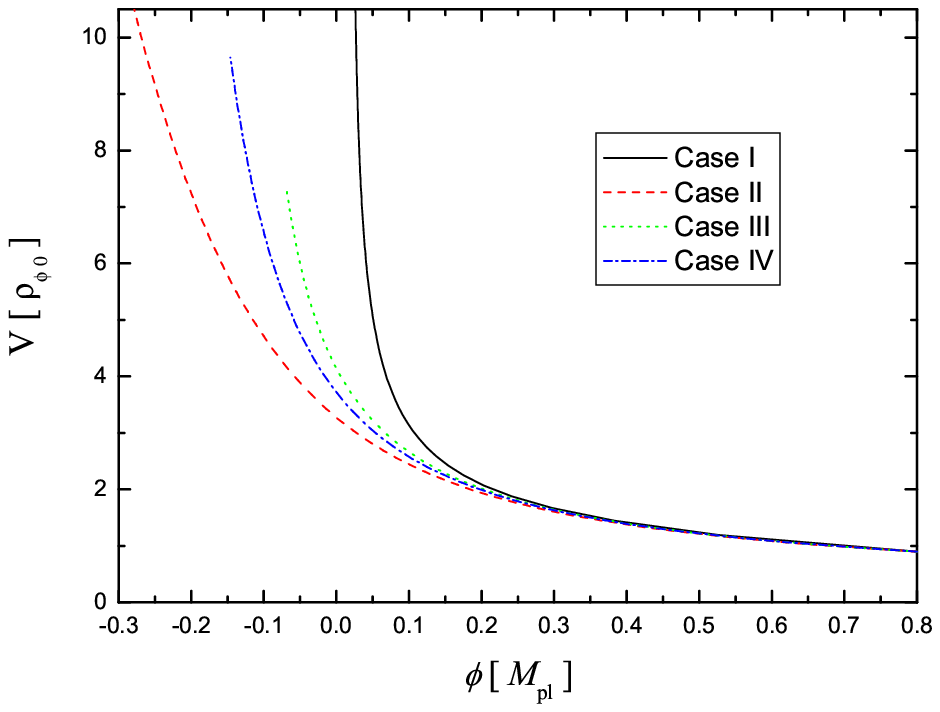}
\caption{Constructed quintessence potentials $V(\phi)$.}
\label{fig:vphi}
\end{center}
\end{figure}

We have numerically evaluated these equations.
Fig.~\ref{fig:rhoz} shows the evolution of the energy density
of the quintessence $\rho_\phi (z)$, where we choose
$w_0=-0.8$, $w_1=0.1$ and $r_0=3/7$. At low redshift, all
models obey the same evolution law, but deviation from this
is clearly visible at redshift $z>1$. Fig.~\ref{fig:vphi}
shows the constructed quintessence potential $V(\phi)$, which
is in the form of a runaway potential.

In the evaluation of these equations, we have also chosen the
initial values of the quintessence field $\tilde \phi_0=0.8$
at the redshift $z=0$ (present).
The value of $\tilde \phi_0$ is chosen for the purpose of
definiteness. If we shift its value, it simply results in the
shift of the value of the scalar field; the potential in
Fig.~\ref{fig:vphi} is shifted horizontally. It has no
influence on the evolution of the universe and the shape of
the quintessence potential. In general, $\phi$ decreases
monotonously as $z$ increases from $-1$, and the potential
increases. This means that the potential decreases as the
universe expands.

As shown in Fig.~\ref{fig:vphi}, the four cases possess the
same asymptotic behavior for the region $0.4 < \phi <0.8$.
This corresponds to low redshift $0<z<1$. We can give the
approximate analytic form of the potential, which is of the
exponential form:
\begin{equation}
\tilde V(\tilde\phi) \simeq \frac{1}{2}(1-w_0)
 \exp\left[-\sqrt{3(1+w_0)(1+r_0)}\;
 (\tilde \phi -\tilde \phi_0)\right].
\label{asym1}
\end{equation}
They differ when $z$ becomes large. In this region, we find
that the field $\phi$ becomes small and the quintessence
potential is in the form of power law for the first case
\begin{equation}
\tilde V(\tilde\phi) \simeq \frac{1}{2}(1-w_0)
 \left(\frac{-w_0\sqrt{3r_0}}{2\sqrt{1+w_0}}\,\tilde\phi\right)
 ^{2(1+w_0)/w_0},
\label{asym2}
\end{equation}
and the potentials are complicated for the last three cases.

Actually it is possible to give an analytic form of the
potential in Case I. We first integrate Eq.~(\ref{phi1}) to
obtain
\begin{equation}
\tilde \phi =\frac{\sqrt{1+w_0}}{\sqrt{3} w_0}\;
\ln \frac{[1+r_0(1+z)^{-3w_0}]^{1/2}-1}{[1+r_0(1+z)^{-3w_0}]^{1/2}+1},
\end{equation}
Solving this for $1+z$ and substituting the result into
Eq.~(\ref{pot1}), we obtain
\begin{equation}
\tilde V(\tilde\phi) = \frac12(1-w_0)\left( \frac{1}{r_0 \sinh^2
(\sqrt{3} w_0\tilde \phi/2\sqrt{1+w_0}\,)} \right)^{-(1+w_0)/w_0}.
\end{equation}
This is of course consistent with the above asymptotic
forms~(\ref{asym1}) and (\ref{asym2}).
This potential was also given in Ref.~\cite{SS}.
We thus confirm that the potential is indeed a runaway type
for $-1<w_0 <0$, and its asymptotic value is zero. This is a
very interesting result since one would obtain such behavior
in general in supersymmetric theories. This kind of potential
is also the one expected for tachyons in unstable D-brane
system in superstring theories~\cite{sen}.


In conclusion, we have developed a method of constructing the
quintessence potential directly from the effective equation of
state function $w_\phi(z)$, which describes the properties of
the dark energy.
Then we have considered four parametrizations of equation of
state parameter and showed that the constructed quintessence
potential takes the form of a runaway type.
The future Supernova/Acceleration Probe with high-redshift
observations~\cite{SNAP},
in combination with the Planck CMB observation~\cite{PLANCK},
will be able to determine the parameters in the dark energy
parametrization to high precision.
By precision mapping of the recent expansion history,
we hope to learn more about the essence of the dark energy.

\section*{Acknowledgements}
This project was in part supported by National Basic Research
Program of China under Grant No. 2003CB716300 and
by NNSFC under Grant No. 90403032.
The work of NO was supported in part by the Grant-in-Aid for
Scientific Research Fund of the JSPS No. 16540250.

\end{document}